%% file: Doug1122.tex
\documentclass[twocolumn]{aastex631}

\usepackage{amsmath}
\usepackage{xspace}
\usepackage{xcolor}
\usepackage{hyperref}
\usepackage{multirow}
\usepackage{tikz-imagelabels}

\newcommand{\VV}{V$^2$\xspace}
\newcommand{\VIDE}{\VV/VIDE\xspace}
\newcommand{\REVOLVER}{\VV/REVOLVER\xspace}

\newcommand{\hMpc}{$h^{-1}$~Mpc\xspace}

\imagelabelset{
    image label distance = 2mm,
    image label back = white,
    image label text = black,
    coordinate label font = \bfseries\scriptsize,
    coordinate label distance = 2mm,
    coordinate label back = white,
    coordinate label text = black,
    annotation font = \normalfont\small,
}

\shorttitle{SDSS DR7 void catalogs}
\shortauthors{Douglass et al.}

\begin{document}

\title{Updated void catalogs of the SDSS DR7 main sample}

\correspondingauthor{Kelly A. Douglass}
\email{kellyadouglass@rochester.edu}

\author[0000-0002-9540-546X]{Kelly A. Douglass}
\affiliation{Department of Physics \& Astronomy, University of Rochester, 500 Wilson Blvd., Rochester, NY  14611}

\author[0000-0001-8101-2836]{Dahlia Veyrat}
\affiliation{Department of Physics \& Astronomy, University of Rochester, 500 Wilson Blvd., Rochester, NY  14611}
\email{dveyrat@ur.rochester.edu}

\author[0000-0001-5537-4710]{Segev BenZvi}
\affiliation{Department of Physics \& Astronomy, University of Rochester, 500 Wilson Blvd., Rochester, NY  14611}
\email{segev.benzvi@rochester.edu}

\begin{abstract}
  We produce several public void catalogs using a volume-limited subsample of 
  the Sloan Digital Sky Survey Data Release 7 (SDSS DR7).  Using new 
  implementations of three different void-finding algorithms, VoidFinder and two 
  ZOBOV-based algorithms (VIDE and REVOLVER), we identify 1163, 531, and 518 
  cosmic voids with radii $> 10$~\hMpc, respectively, out to a redshift of 
  $z = 0.114$ assuming a Planck 2018 cosmology, and 1184, 535, and 519 cosmic 
  voids assuming a WMAP5 cosmology.  We compute effective radii and centers for 
  all voids and find none with an effective radius $> 54$~\hMpc.  The median 
  void effective radius is 15--19~\hMpc for all three algorithms.  We extract 
  and discuss several properties of the void populations, including radial 
  density profiles, the volume fraction of the catalog contained within voids, 
  and the fraction of galaxies contained within voids.  Using 64 mock galaxy 
  catalogs created from the Horizon Run 4 $N$-body simulation, we compare 
  simulated and observed void properties and find good agreement between the 
  SDSS~DR7 and mock catalog results.
\end{abstract}

\section{Introduction}

Galaxy redshift surveys have revealed a web-like distribution of galaxies in the 
observable universe \citep{Bond96}.  The cosmic web exhibits dense clusters 
connected by extended filaments separated by vast regions nearly empty of 
galaxies, known as voids.  \cite{Rood88} reviewed the paradigm shift that 
started in the mid-1970s as galaxy redshift surveys began to develop in their 
depth, density, and sky coverage, enabling more detailed study of the 
three-dimensional spatial distribution of galaxies and the impact that this had 
on void studies.  Early studies by \cite{Joeveer78}, \cite{Gregory78}, and 
\cite{Kirshner81} indicated the presence of large, underdense regions in the 
galaxy distribution, and the Center for Astrophysics Redshift Survey 
\citep{Huchra83} and its extension to $m_B = 15.5$ 
\citep{deLapparent86, Geller89} revealed that voids dominate the large-scale 
structure of the universe.  The voids likely evolved from primordial 
underdensities in the early universe and are a ubiquitous feature of 
gravitational instabilities and the growth of structure 
\citep[and references therein]{vandeWeygaert11}.

Because voids possess much lower densities relative to clusters and filaments, 
their gravitational evolution remains in the linear regime through much of their 
formation, and at later times they become dominated by dark energy earlier than 
other regions of the universe \citep{Goldberg04}.  As discussed in the review by 
\cite{vandeWeygaert11}, the unique properties of void environments make them 
important probes of cosmology and galaxy formation.  For example, voids have 
been used in Alcock-Paczy{\'n}ski tests \citep{Lavaux12, Sutter12b, Sutter14b, 
Hamaus16, Mao17a, Nadathur19, Hamaus20}, to constrain the dark energy equation 
of state \citep[e.g.,][]{Lavaux12, Pisani15, Verza19}, and to measure the scale 
of baryon acoustic oscillations \citep[e.g.,][]{Kitaura16, Liang16, Nadathur19, 
Zhao20, Zhao21}.  In addition, galaxies are known to evolve differently in voids 
compared to galaxies in denser environments, collapsing earlier and forming 
stars later.

Searches for voids in the distribution of galaxies have historically been based 
on one of two strategies: either by locating relatively empty spherical regions 
in the galaxy distribution, or using a watershed algorithm to link low-density 
regions.  The most widely-used void catalogs produced to date are based on the 
7th data release of the Sloan Digital Sky Survey \citep[SDSS~DR7;][]{SDSSdr7}, 
which is complete to a Petrosian $r$-band magnitude $m_r = 17.77$.  The first 
SDSS~DR7 void catalog, published by \cite{Pan12}, was constructed from a union 
of relatively empty spheres and has been used to study void structure 
\citep{Alpaslan14, Lee15, Shim15}, the properties of galaxies in voids 
\citep{Hoyle12, Moorman14, Moorman15, Liu15, Penny15, FraserMcKelvie16, 
Moorman16, Douglass17a, Douglass17b, Douglass18, Douglass19, Miraghaei20}, the 
intergalactic medium in voids \citep{Tejos12}, and for cosmological analyses 
\citep{Ilic13, Planck14}.  The void catalog of \cite{Sutter12a} is based on the 
watershed algorithm and has been used to study the properties of void galaxies 
\citep{Furniss15, Trevisan21} and for cosmological analyses \citep{Sutter12b, 
Ilic13, Hamaus14a, Sutter14b, Pisani14, Planck14, Cai14, Melchior14, Chen15, 
Dai15, Chantavat17}.

In this work, we present updated void catalogs based on the SDSS~DR7 galaxy 
sample from the NASA-Sloan Atlas \citep[NSA;][]{Blanton11}.  We have generated 
catalogs using the two main void-finding algorithms, which have been implemented 
in the Void Analysis Software Toolkit \citep[VAST;][]{Douglass22}.  These 
implementations include improved handling of survey boundaries and increased 
computational efficiency for void finding. This is the first time that void 
catalogs have been generated using complementary algorithms on the same 
volume-limited\footnote{Previous void catalogs produced in SDSS~DR7 did not use 
the same volume-limited sample.} galaxy survey.  We use the SDSS~DR7 catalog 
because it remains the most complete galaxy catalog of the nearby universe 
produced to date.

The paper is structured as follows.  In Section~\ref{sec:data} we describe the 
SDSS~DR7 data sample, the cuts used to produce a volume-limited catalog, and the 
simulation used to produce mock catalogs.  We describe the void-finding 
algorithms in Section~\ref{sec:vast} and the resulting void catalogs in 
Section~\ref{sec:catalogs}.  In Section~\ref{sec:discussion}, we compare our 
catalogs with previous results.


\section{Data and Simulations}\label{sec:data}

\subsection{Data: SDSS DR7}

Released in 2009, the SDSS~DR7 \citep{SDSSdr7} is a wide-field multiband imaging 
and spectroscopic survey conducted on the 2.5~m telescope at the Apache Point 
Observatory in New Mexico.  Photometric data in the SDSS five-band system 
\citep{Fukugita96, Gunn98} was collected using a drift scanning technique to 
image approximately one quarter of the night sky.  Follow-up spectroscopy was 
performed on all galaxies with a Petrosian $r$-band magnitude $m_r < 17.77$ 
\citep{Lupton01, Strauss02} with two double fiber-fed spectrometers and fiber 
plug plates with a minimum fiber separation of 55''.  The observed wavelength 
range for the SDSS~DR7 spectra is 3800--9200\AA~with a resolution 
$\lambda/\Delta \lambda \sim 1800$ \citep{Blanton03}.

We use version 1.0.1 of the NASA-Sloan Atlas \citep[NSA;][]{Blanton11}, which 
contains 641,409 galaxies observed in SDSS~DR7.  We construct a volume-limited 
sample from the NSA by requiring all objects to have an absolute $r$-band 
magnitude $M_r \leq -20$ and redshift $z \leq 0.114$.  This is necessary to 
remove the natural decrease in matter tracer density with increasing redshift 
present in magnitude-limited surveys, as this would introduce an artificial 
relationship between void radius and redshift.  Our luminosity limit is selected 
to retain all galaxies brighter than $L_*$, which have been shown to 
sufficiently trace the cosmic web structure \citep{Pan12}.  The redshift limit 
corresponds to the distance at which these $M_r = -20$ objects are 
spectroscopically observed by SDSS~DR7.  Within the main SDSS~DR7 footprint, 
194,125 galaxies are in the volume-limited sample.

\subsection{Mock Catalogs: Horizon Run 4 Simulations}

We create 64 mock galaxy catalogs generated from the Horizon Run~4 Simulation 
\citep{Kim15} to estimate the variances of the void spectra and density 
profiles.  Horizon Run~4 is an $N$-body simulation generated from $6300^3$ 
particles in a cubic box with side length 3150~\hMpc using the WMAP 5-year 
$\Lambda$CDM cosmology: $\Omega_m = 0.258$, $\Omega_\Lambda = 0.742$, 
$h = 0.719$, $\sigma_8 = 0.796$, and $n_s = 0.963$ \citep{Dunkley09}.  As part 
of Horizon Run~4, \cite{Hong16} produced a mock galaxy catalog, calibrating it 
to match the properties of galaxy groups in volume-limited SDSS galaxy catalogs.  
The resulting mock catalog is divided into 64 cubic equal-volume subsamples.

To replicate the angular and redshift distribution of galaxies in SDSS~DR7, we 
create a mock catalog from each Horizon Run~4 subsample by applying cuts in 
right ascension, declination, and redshift relative to an observer positioned at 
the center of the cubic volume.  Further, the mock galaxy catalogs provide a 
mass-like quantity computed from the friends-of-friends halo mass 
\citep{Hong16}, which can be used as a proxy for luminosity.  By applying a cut 
on this quantity, we match the redshift selection function of the volume-limited 
SDSS~DR7 catalog in each of the 64 mocks.


\section{The Void Analysis Software Toolkit}\label{sec:vast}

We construct voids using the Void Analysis Software Toolkit\footnote{VAST is 
available for download at \url{https://github.com/desi-ur/vast}.} 
\citep[VAST;][]{Douglass22}.  VAST provides Python-only implementations of two 
popular classes of void-finding algorithms: VoidFinder, which identifies voids 
by growing and merging nearly empty spherical regions in a galaxy catalog; and 
\VV, which uses a watershed algorithm to link low-density cells in a Voronoi 
tesselation of the galaxy catalog.

\subsection{VoidFinder}\label{sec:VF}

Originally described by \cite{ElAd97} and implemented by \cite{Hoyle02}, 
VoidFinder identifies apparent dynamically distinct voids by growing empty 
spheres within a galaxy distribution and then merging the spheres to form voids.  
VoidFinder first locates and removes all isolated, or field, galaxies, defined 
by the distance to their third-nearest neighbor, and then places the remaining 
non-isolated, or wall, galaxies on a 3D grid.  A sphere is then grown from each 
empty cell in the grid until it is bounded by four galaxies.  To keep voids from 
extending beyond the survey edges, a sphere is removed if more than 10\% of its 
volume lies outside of the survey. 

To extract apparent dynamically-distinct voids from the resulting list of 
spheres, the spheres are sorted by radius, and the largest is identified as a 
maximal sphere --- the largest sphere that can fit inside a given void region.  
Each subsequent sphere with a radius larger than 10~\hMpc is also labeled as a 
maximal sphere if it does not overlap any already-identified maximal sphere by 
more than 10\% of its volume.  To refine the volume and profile of the void 
regions, all spheres not identified as maximal spheres are merged with a maximal 
sphere if they overlap exactly one maximal sphere by at least 50\% of their 
volume.  A void is therefore defined as a union of spheres: one maximal sphere, 
plus some number of smaller spheres.

\subsection{Voronoi Voids (\VV)}\label{sec:V2}

Voronoi Voids, or \VV, is a Python-only implementation based on the ZOBOV 
algorithm \citep{Neyrinck08}.  It first produces a 3D Voronoi tessellation of 
the input catalog to separate each galaxy into a primitive cell. Each cell’s 
volume is computed and used to estimate the local density in 3D. The Voronoi 
cells are subsequently combined into groups or ``zones'' using watershed 
segmentation.  Each cell is put into the same zone as its least-dense neighbor, 
and any cell less dense than any of its neighboring cells --- a local density 
minimum --- is identified as its zone’s central cell.

Zones are merged to form voids by first identifying the least-dense pair of 
adjacent cells between two zones. The set of all zones is partitioned into a 
subset of voids linked together by zones of equal or lower density, called the 
``linking density.''  The algorithm loops over linking density values to produce 
the voids.  Because this zone-linking procedure produces a hierarchy of voids 
ranging from individual small zones to one huge void encompassing the survey 
volume, a pruning step is used to fix the final number of voids.  This can 
proceed either by setting a user-defined maximum zone linking density 
\citep[VIDE pruning;][]{Sutter15}, or by estimating the probability that any 
given void is due to catalog shot noise and removing the voids most likely 
produced by shot noise \citep[ZOBOV pruning;][]{Neyrinck08}.  Alternatively, one 
can simply define voids as single zones with volume above some threshold, such 
as the median volume of all zones \citep[REVOLVER pruning;][similar to VIDE 
pruning with an arbitrarily small zone-linking density and a minimum radius 
cut]{Nadathur19}.  The density threshold for a given pruning method can also be 
determined by computing the effective radius of an equivalent spherical void,
\begin{equation}\label{eqn:reff}
    R_\text{eff} = \left( \frac{3V}{4\pi} \right)^{1/3}
\end{equation}
where $V$ is the void volume, and then selecting $R_\text{eff}$ greater than 
some user-defined threshold.  To compute the center of each void, \VV uses a sum 
over the positions of the constituent galaxies, weighted according to their 
Voronoi volumes.

The use of a Voronoi tessellation assigns arbitrarily large volumes to cells 
near the boundary of the survey, impacting the calculation $R_\text{eff}$ in 
Eqn.~\eqref{eqn:reff}.  To account for this effect, \VV assigns a volume of 0 to 
any cell that extends outside the edge of a survey.  Effectively, the edge is 
treated as an infinitely dense wall surrounding the survey through which zones 
cannot penetrate.

\section{SDSS DR7 void catalogs}\label{sec:catalogs}

\begin{figure}
  \centering
  \begin{annotationimage}{width=0.49\textwidth}{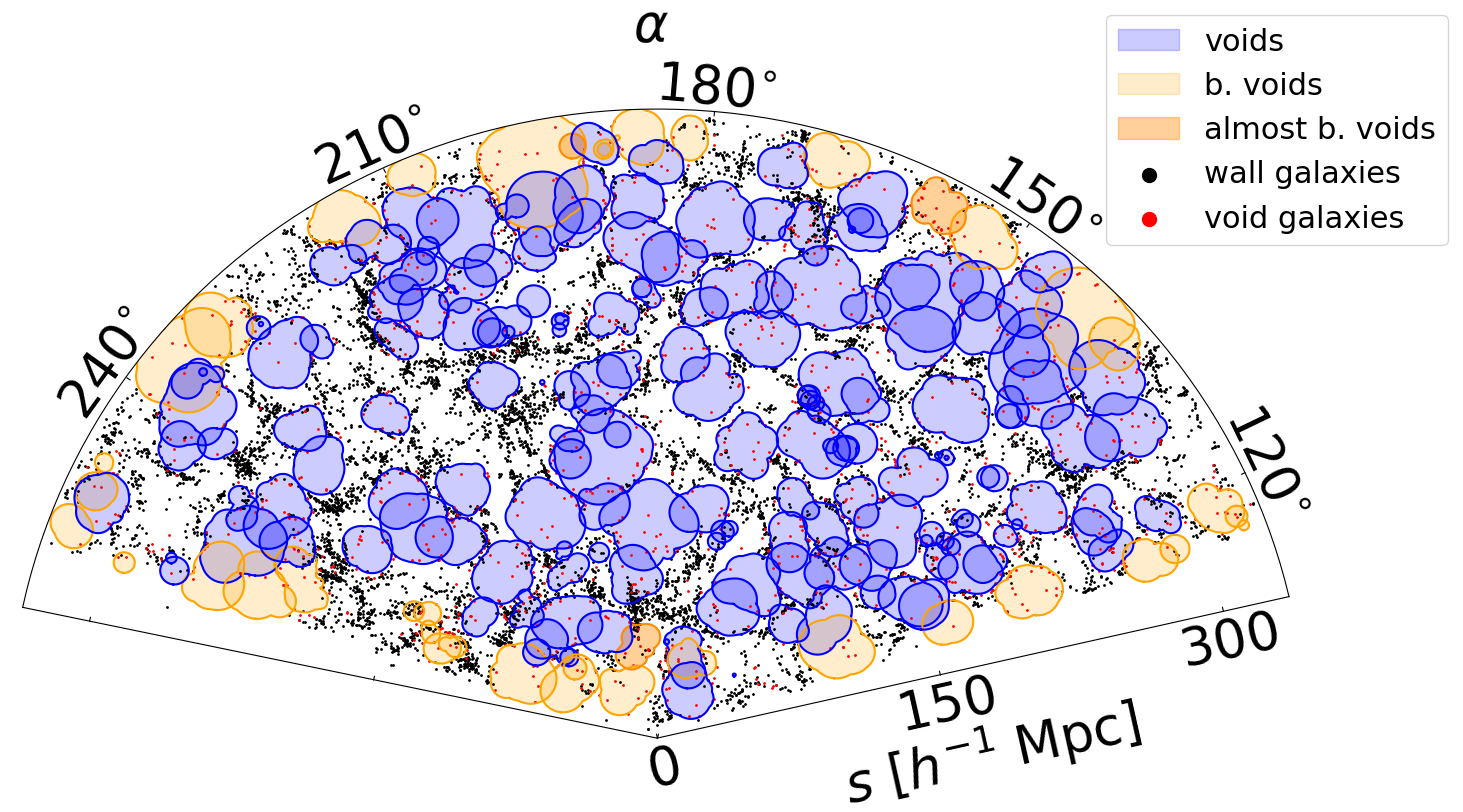}
    \draw[image label = {{VoidFinder} at north west}];
  \end{annotationimage}
  \begin{annotationimage}{width=0.49\textwidth}{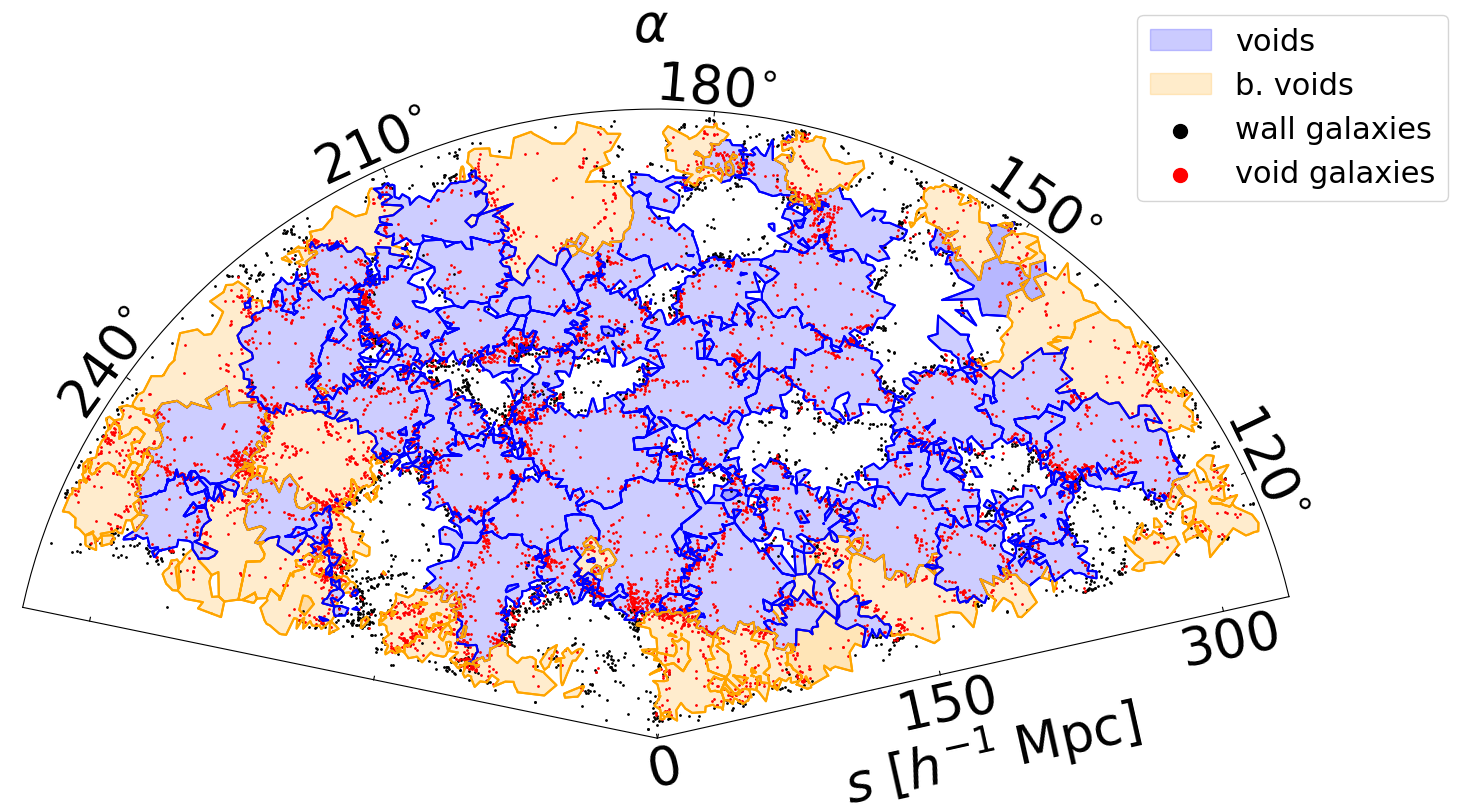}
    \draw[image label = {{\VIDE} at north west}];
  \end{annotationimage}
  \begin{annotationimage}{width=0.49\textwidth}{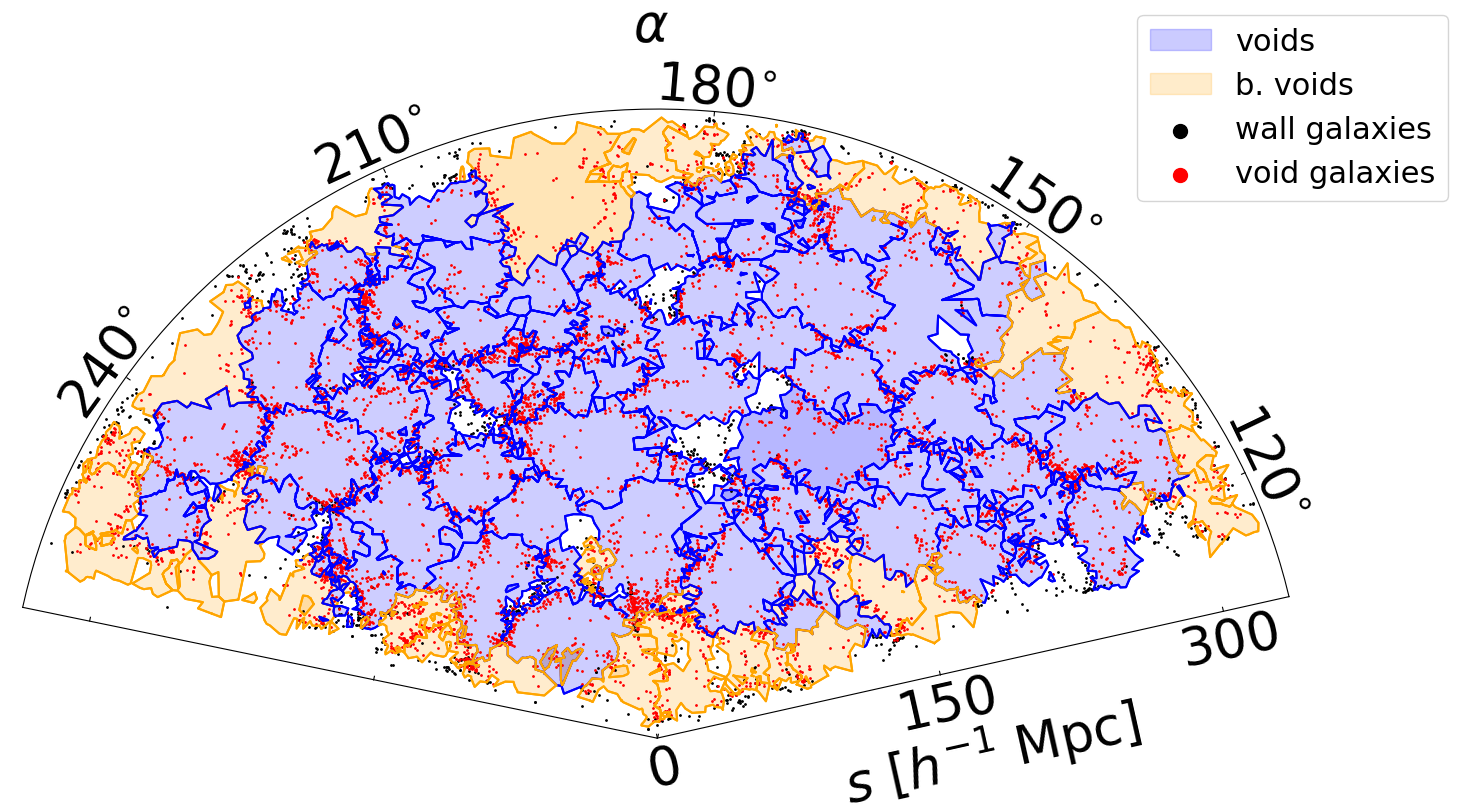}
    \draw[image label = {{\REVOLVER} at north west}];
  \end{annotationimage}
  \caption{ A 10~\hMpc thick slice of SDSS~DR7 centered at a declination of 
  $25\deg$ assuming Planck 2018 cosmology.  The locations of the galaxies which 
  fall within a void are shown in red, while those which do not are shown in 
  black.  The intersection of all voids with the center of this slice are shown 
  in the shaded blue (orange for boundary voids) regions, for VoidFinder (top), 
  \VIDE (middle), and \REVOLVER (bottom).}
  \label{fig:dec_slices}
\end{figure}

We present void catalogs found with VoidFinder, \VV with VIDE pruning (\VIDE), 
and \VV with REVOLVER pruning (\REVOLVER) assuming two flat $\Lambda$CDM 
cosmologies: WMAP5 \citep[$\Omega_M = 0.258$;][]{Dunkley09} and Planck 2018 
\citep[$\Omega_M = 0.315$;][]{Planck18}, and assuming $H_0 = 100h$~km/s/Mpc.  
Each void-finding algorithm produces a different void catalog for the same 
volume-limited galaxy sample.  Examples of the void locations relative to the 
galaxies are shown in Figure~\ref{fig:dec_slices} for VoidFinder (top), \VIDE 
(middle), and \REVOLVER (bottom).  The galaxies that fall within a void region 
defined by the corresponding void catalog are colored in red, while those that 
do not are shown in black.  The three void catalogs identify approximately the 
same regions as voids, but the details of their borders, including exactly where 
the void ends, is very different between VoidFinder and \VV.  The effect that 
this has on the inferred properties of void galaxies is discussed in Zaidouni et 
al. (in prep.).

\input{VF_maximals}

\vspace{-0.35in}

Our void catalogs are publicly available at 
\href{https://doi.org/10.5281/zenodo.5834786}{https://doi.org/10.5281/zenodo.5834786}.  
Both algorithms produce output files in \texttt{ASCII} format.  The file 
contents are different due to the differences in the void-finding algorithms 
and are described below.

\input{VF_holes}

\vspace{-0.15in}

The VoidFinder void catalog consists of two files, one containing the maximal 
spheres and the other containing a list of all of the spheres that are part of a 
void.  The maximal sphere file contains the coordinates of the centers of the 
maximal spheres (in both Cartesian and sky coordinates), along with their radii, 
unique identifiers, the effective radii of the voids, and whether the void sits 
close to the edge of the survey.  We use the following transformations to 
convert from sky coordinates ($\alpha$, $\delta$) to Cartesian coordinates:
\begin{eqnarray}\label{eq:coord_transforms}
    x &=& s\cos(\alpha)\cos(\delta)\\
    y &=& s\sin(\alpha)\cos(\delta)\\
    z &=& s\sin(\delta)
\end{eqnarray}
where $s$ is the comoving distance to the galaxy using one of the two flat 
$\Lambda$CDM cosmologies specified earlier.  The first few lines of this file 
are shown in Table~\ref{tab:vffile1}.  The file listing all of the spheres 
(including the maximal spheres) contains each sphere's center in Cartesian 
coordinates, its radius, and the unique identifier corresponding to its void's 
maximal sphere.  A portion of this file can be found in Table~\ref{tab:vffile2}.  
The union of all spheres with the same unique identifier define one void.

We provide two \VV void catalogs, one with VIDE pruning \citep[maximum zone 
linking density of $0.2\overline{n}$, as described in][]{Sutter15} and a minimum 
$R_\text{eff}$ of 10~\hMpc, and the other with REVOLVER pruning.  The minimum 
radius limit for the \VIDE catalog is chosen to closely match the minimum 
maximal sphere radius of 10~\hMpc in VoidFinder's void catalog.  Each of the \VV 
void catalogs consists of three files, one containing the details of the voids, 
one describing the connection between the zones and voids, and one describing 
the connection between the galaxies and zones.  The void file contains the 
Cartesian coordinates, redshift, and sky coordinates of each void's 
volume-weighted center\footnote{The volume-weighted center the average position 
of all galaxies in the voids, weighted by the Voronoi volume of each galaxy.}, 
along with each void's effective radius (Eqn.~\ref{eqn:reff}), the principal 
axes of the ellipsoid fitted to each void, the total surface area of the void, 
and the surface area of the void shared with a boundary galaxy cell.  The zone 
file contains each zone's unique identifier, the smallest void to which it 
belongs, and the largest void to which it belongs.  The galaxy file contains 
each galaxy's unique identifier (from the input galaxy catalog), the zone to 
which it belongs, the number of adjacent cells between its Voronoi cell and the 
edge of the zone to which it belongs, and flags to identify whether it is at the 
edge of or outside the survey mask.  Excerpts from these files are shown in 
Tables~\ref{tab:v2file1}--\ref{tab:v2file3}.

\subsection{Defining boundary voids}\label{sec:boundaries}

\begin{figure}
  \includegraphics[width=0.5\textwidth]{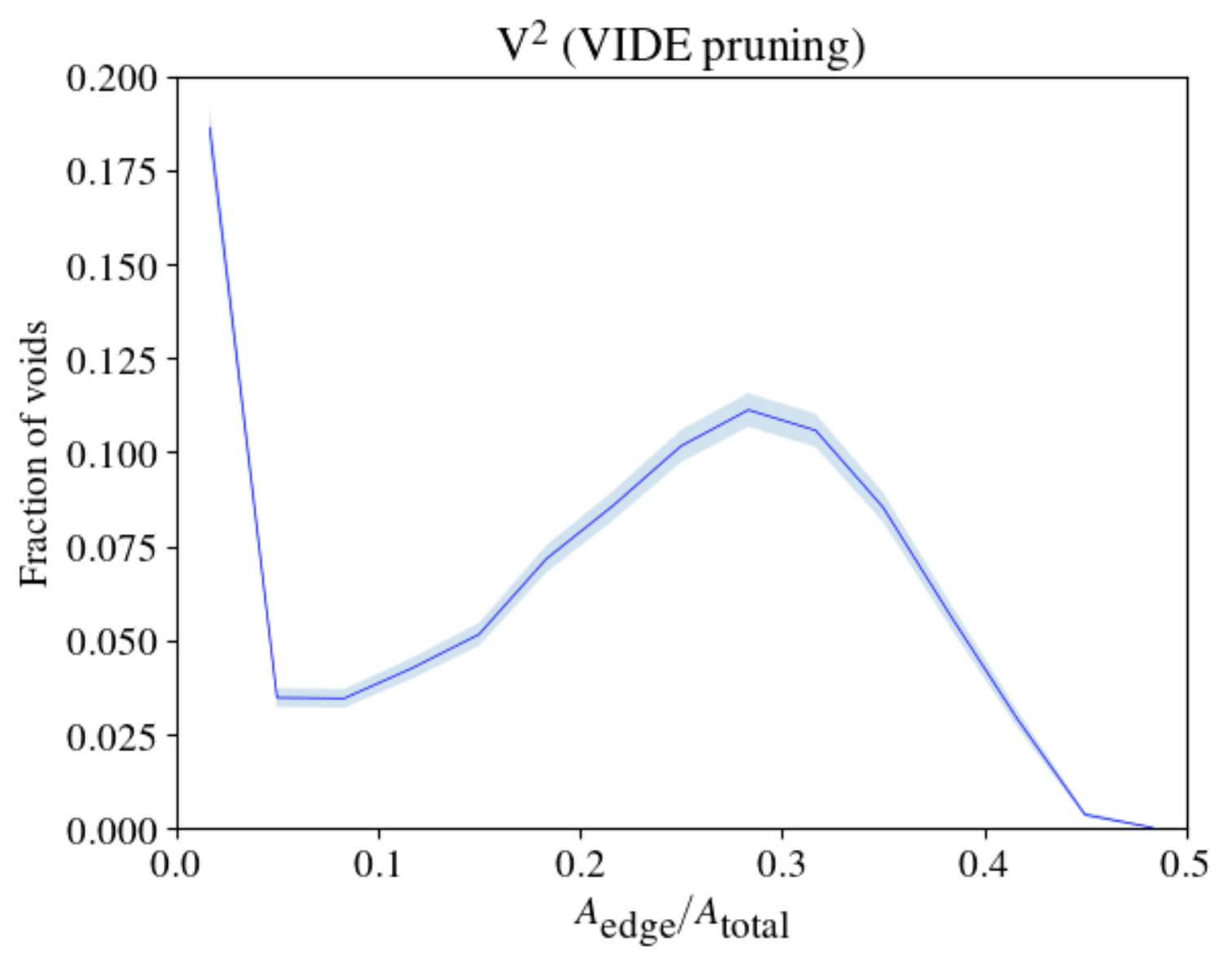}
  \caption{Distribution of the ratio of the surface area of \VV voids shared 
  with boundary cells ($A_\text{edge}$) to the total surface area of each void 
  ($A_\text{total}$), for all voids with $A_\text{edge} > 0$.}
  \label{fig:V2_boundary}
\end{figure}

Voids found along the edges of the galaxy survey can be subject to boundary 
effects, as there is no information for which the voids can be constrained 
beyond the survey boundary.  For each void in all catalogs, we include 
information to indicate if the void is potentially affected by the finite volume 
of the survey, so that the user can easily filter these out if their analysis is 
sensitive to these boundary conditions.

In the case of VoidFinder, we define boundary voids as those with any part of 
their surface extending beyond the survey mask.  These voids are flagged with a 
value of 1 in the \texttt{edge} column of the maximal spheres file.  In our 
SDSS~DR7 void catalog presented here, this includes 383 voids, or $\sim$33\% of 
the voids in the catalog.  These voids are shown in orange in the top panel of 
Figure~\ref{fig:dec_slices}.  We also identify voids that have at least one 
sphere with a center within 10\hMpc of the survey mask; these are flagged with a 
value of 2 in the \texttt{edge} column.

For \VV, we define boundary voids as those with a significant fraction of their 
surface area shared with a boundary galaxy's Voronoi cell (described in 
Sec.~\ref{sec:V2}.  Included in the \VV void files are columns containing the 
total surface area of the voids and the surface area of the voids shared with 
these boundary cells.  To determine which \VV voids are significantly impacted 
by the survey bounds, we examine the distribution of the ratio of the surface 
area of \VV voids shared with boundary cells, $A_\text{edge}$, to the total 
surface area of each void, $A_\text{total}$.  As shown in 
Fig.~\ref{fig:V2_boundary}, we find two distinct populations in the distribution 
between voids with surface area shared with boundary cells less than and greater 
than $\sim$10\% of the total surface area of the void.  For studies that are 
sensitive to an inaccurate $R_\text{eff}$ (e.g., the void spectrum) shown below, 
we exclude voids with $A_\text{edge}/A_\text{total} > 0.1$; this includes 234 
voids ($\sim$44\%) in the \VIDE and 216 voids ($\sim$42\%) in the \REVOLVER 
SDSS~DR7 void catalogs.  These voids are shown in orange in 
Figure~\ref{fig:dec_slices}.  We include the areas in the catalogs so that the 
user has the freedom to be as restrictive as they find necessary for their 
analysis.

\subsection{Properties of the void catalogs}\label{sec:prop}

We find 1163 voids in SDSS~DR7 with VoidFinder, 531 voids with \VIDE, and 518 
voids with \REVOLVER assuming Planck 2018 cosmology, and 1184, 534, and 518 
voids assuming WMAP5 cosmology, respectively.  The voids found by VoidFinder 
have a median effective radius $R_\text{eff} = 15.5\pm0.1$~\hMpc, and contain 
$\sim$61\% of the survey volume and $\sim$18\% of the galaxies in the 
volume-limited DR7 catalog.  (The survey volume contained within voids is 
estimated by counting up the fraction of test particles placed on a 1~\hMpc grid 
that fall within one of the VoidFinder void spheres.)  With \VIDE, voids have a 
median $R_\text{eff} \simeq 17$~\hMpc ($\simeq 19.5$~\hMpc in the case of 
\REVOLVER), and contain 63--68\% (86--90\%) of the survey volume and 
$\sim$64\% ($\sim$82\%) of the volume-limited galaxies.  While all three 
algorithms resulted in similar median effective radii, the $R_\text{eff}$ of the 
largest void differs significantly, ranging from 28.7~\hMpc with VoidFinder to 
53.3~\hMpc with \VIDE.  The catalog statistics are summarized in 
Table~\ref{tab:data_table}.

There is a significant difference in the fraction of galaxies contained within 
voids found by VoidFinder and voids found by \VV.  While the void-growing step 
in VoidFinder explicitly stops once wall galaxies are encountered, \VV voids 
extend to the peaks of the density maxima surrounding them.  Thus, many galaxies 
in overdense regions are incorrectly classified as void galaxies by \VV, as can 
be seen by the significant fraction of void galaxies in extremely dense regions 
of the galaxy distribution shown in the middle and lower panels of 
Figure~\ref{fig:dec_slices}.  \VV is not ideal for isolating the void 
environment, a crucial detail for void-galaxy classification \citep[Zaidouni et 
al., in prep.]{Florez21}.  However, \VV is useful for studying galaxy properties 
as a function of large-scale environment by using the volumes of the Voronoi 
cells as a proxy for local density \citep[e.g.,][]{Goh19, Habouzit20, 
Paranjape20, Jamieson21}.

\begin{figure}
  \centering
  \includegraphics[width=0.49\textwidth]{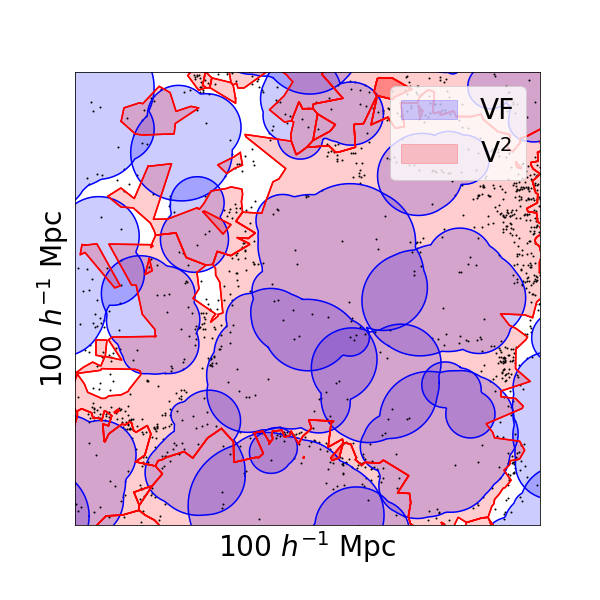}
  \caption{Example volume overlap between two void catalogs: VoidFinder (blue) 
  and \VV (red).  Within the survey volume, we calculate the fraction of the 
  space that both catalogs classify as a void (purple).}
  \label{fig:overlap}
\end{figure}

\input{V2_voids_skinny}

\input{V2_zones}

\input{V2_galaxies}

\vspace{-5em}

Because VoidFinder and \VV define voids differently, we study their application 
to the survey volume so that we can quantitatively compare their void catalogs.  
As demonstrated in Figure~\ref{fig:overlap}, we compute the percentage of the 
survey volume that two catalogs both consider to be a void (shaded in purple in 
Figure~\ref{fig:overlap}), and we compute the percentage of the volume-limited 
galaxy catalog that two algorithms both classify to be void galaxies.  In other 
words, for any two void catalogs $A$ and $B$, we compute the volume of $A\cap B$ 
and report the fraction of the survey inside the volume $A\cap B$.  We also 
compute the percentage of galaxies inside the volume $A\cap B$.  If the 
fractional survey volume in voids in catalog $A$ from Table~\ref{tab:data_table} 
is close to the fractional volume in voids in $A\cap B$ 
(Table~\ref{tab:overlap}), it implies that the void volume in catalog $B$ 
contains most of the void volume in catalog $A$.

For example, consider \REVOLVER.  The fractional survey volumes in voids for 
\REVOLVER $\cap$ VoidFinder and \REVOLVER $\cap$ VIDE are within a few percent 
of the fractional survey volumes identified as voids with VoidFinder and \VIDE 
alone.  This is unsurprising: since \REVOLVER identifies $\sim$90\% of the 
survey volume to be within voids, the \REVOLVER voids tend to contain the voids 
identified by VoidFinder and/or \VIDE.  These overlaps can also be seen in 
Figure~\ref{fig:dec_slices}, where it is visually clear that the \REVOLVER voids 
(red areas) tend to contain the VoidFinder voids (blue areas).

In contrast, the percent survey volume identified commonly as void regions with 
both VoidFinder and \VIDE ($\sim$40\%) reveals significant differences between 
these two catalogs.  VoidFinder identifies $\sim$61\% of the survey volume as a 
void, indicating that $\sim$20\% of the survey volume identified as void by 
VoidFinder is classified as wall by \VIDE.  This can be seen in 
Figure~\ref{fig:dec_slices}, where there are apparent underdense regions 
classified as a void by VoidFinder but not included in either \VV catalog.

For comparison, we also compute the fraction of void volume overlap and void 
galaxy overlap using the 64 mock catalogs and report them at the bottom of 
Tables~\ref{tab:data_table} and \ref{tab:overlap}.  Overall there is good 
agreement with data, which tend to have slightly fewer voids than the mocks.  
The large variation in maximum $R_\text{eff}$ for \VIDE is likely a 
manifestation of the large relative uncertainty in the frequency of the 
zone-linking process that produces the largest voids in the catalog.

\subsection{Void spectrum}\label{sec:spec}

\begin{figure*}
  \includegraphics[width=\textwidth]{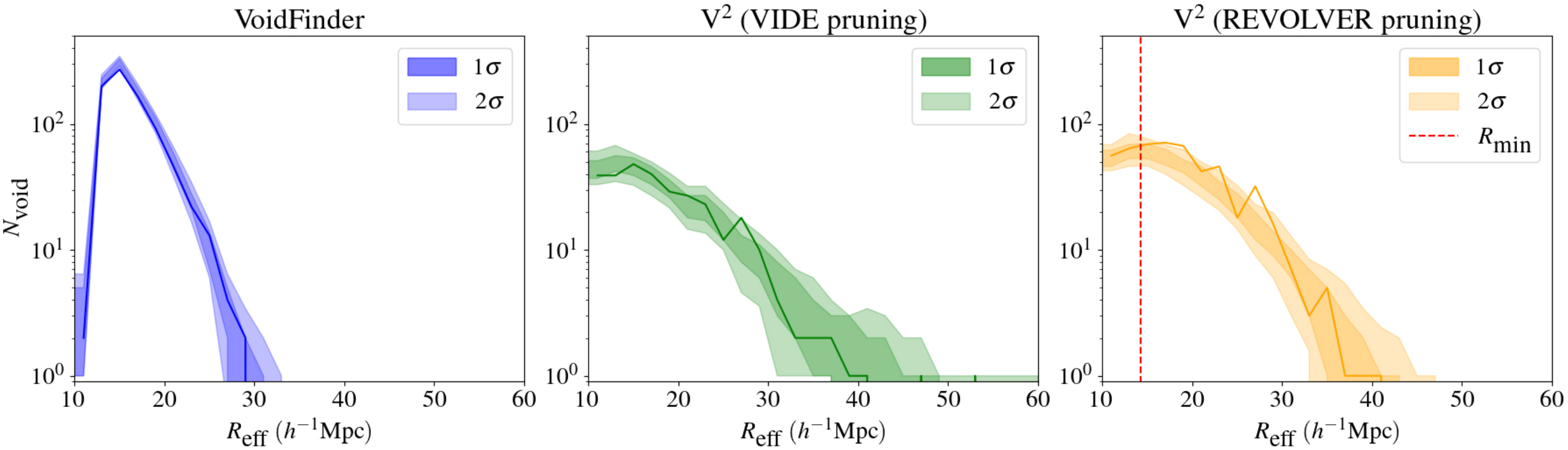}
  \caption{Distribution of voids' effective radii for the VoidFinder (left), 
  \VIDE (center), and \REVOLVER (right) SDSS DR7 void catalogs (solid lines) 
  generated assuming WMAP5 cosmology.  The shaded uncertainties were computed 
  using 64 DR7 mock catalogs produced from the Horizon Run 4 simulation 
  \citep{Kim15}.  VoidFinder voids are more numerous and generally smaller due 
  to how \VV aggregates disconnected voids (see Section~\ref{sec:spec} for 
  discussion).  We extend the \REVOLVER distribution down to 10~\hMpc so that it 
  is consistent with both VoidFinder and \VIDE.}
  \label{fig:vspec}
\end{figure*}

The distributions of void effective radii are shown in Figure~\ref{fig:vspec} 
for VoidFinder (left), \VIDE (middle), and \REVOLVER (right).  Boundary voids, 
as defined in Sec.~\ref{sec:boundaries}, are excluded from these spectra, as 
their radii are likely affected by the finite volume of SDSS~DR7.  Because the 
maximal spheres in VoidFinder are limited to a 10~\hMpc minimum radius as 
opposed to the 10~\hMpc minimum effective radius cut on the \VIDE, the peak in 
the void spectrum for VoidFinder occurs at a larger radius than 10~\hMpc.  
Additionally, the VoidFinder distribution drops off at $\sim$30~\hMpc, while for 
the \VV algorithms this occurs at 40--50~\hMpc.  This is a result of VoidFinder 
identifying apparent dynamically distinct voids, which evolve independently and 
are smaller and more numerous than the voids found by \VV.

Our distributions of the void effective radii agree reasonably well with the 
catalogs of \cite{Pan12} and \cite{Sutter12a}.  The most significant differences 
occur between our VoidFinder catalog and that of \cite{Pan12} and are a result 
of our improved treatment of the survey boundaries and sphere merging in our 
implementation of VoidFinder (see Section~\ref{sec:discussion} for details).

\subsection{Void density profile}

\begin{figure*}
  \includegraphics[width=\textwidth]{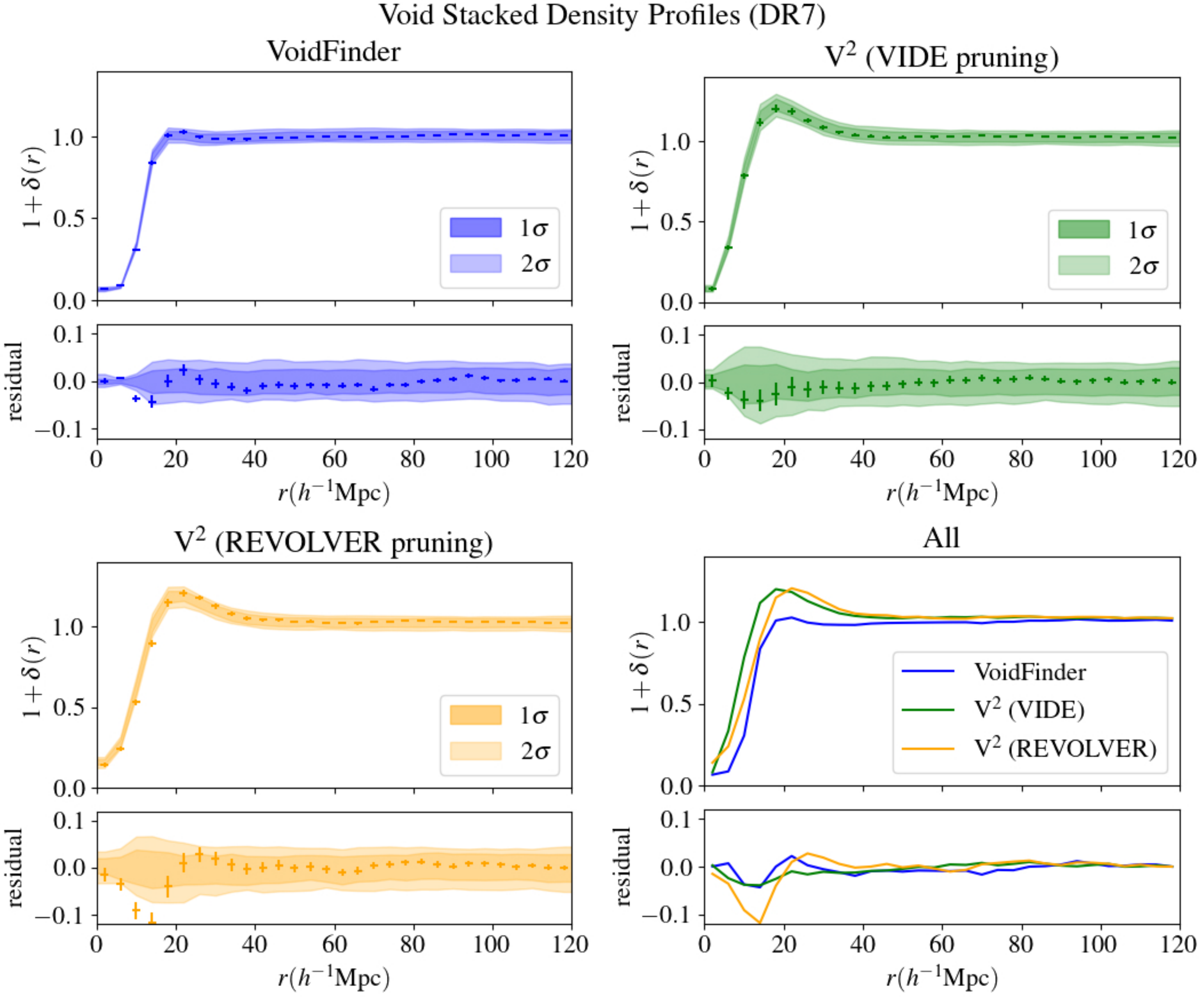}
  \caption{Stacked radial void density profiles, measured in concentric 
  spherical shells around void centers generated assuming the WMAP5 cosmology.  
  We estimate the profiles using the average density of all galaxies, measured 
  from the centers of voids.  $1 + \delta = 1$ corresponds to the mean survey 
  density, while $1 + \delta = 0$ corresponds to a density of 0.}
  \label{fig:vdel}
\end{figure*}

Gravitational theory predicts that void boundaries form from shell crossing, 
where the void interior expands faster than the collapsing denser regions 
\citep{Sheth04}.  As a result, we expect voids to be surrounded by regions of 
higher density than average.  While voids tend to exhibit an underdensity of 
galaxies near their centers, the radial density profiles of individual voids 
vary considerably within the population.  For this reason, it is useful to 
define a stacked void density profile, where the density $n$ in each radial bin 
is computed from a sum over all voids of the number of galaxies in that bin 
normalized by the volume of that bin that is within the survey.  The spherically 
averaged stacked density profile $\delta(r)$ is then defined as 
\begin{equation}
    \delta(r) = \frac{n(r)}{\bar{n}}-1,
\end{equation}
where $\bar{n}$ is the average density of the survey.

The distribution over $1 + \delta(r)$ is shown in Figure~\ref{fig:vdel} for the 
VoidFinder (top left), \VV using \VIDE (top right), and \REVOLVER (bottom left) 
SDSS~DR7 void catalogs.  As expected, the density profiles exhibit the expected 
behavior of approaching 1 at large $r$.

\begin{figure*}
  \includegraphics[width=\textwidth]{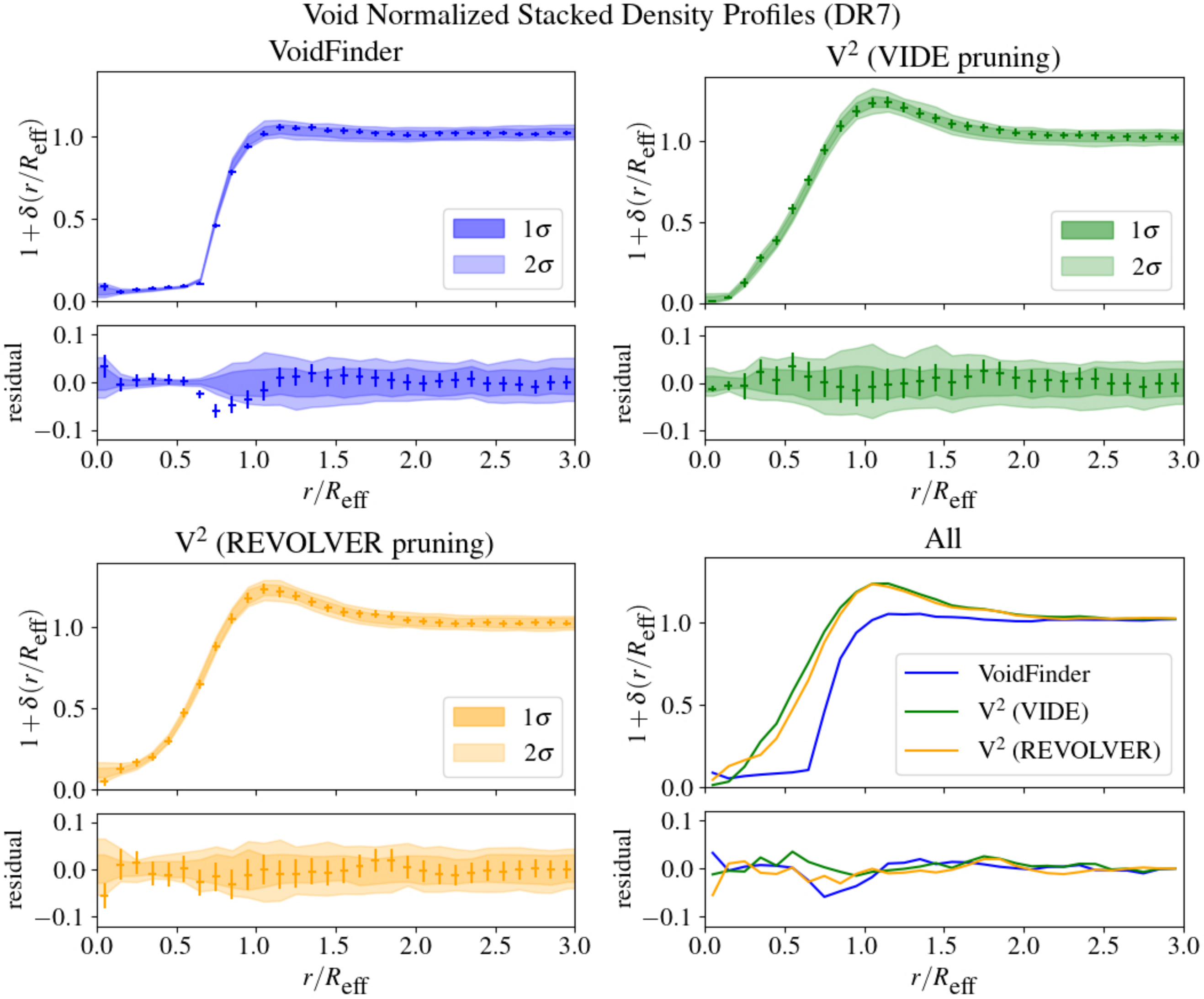}
  \caption{Stacked normalized radial void density profiles assuming WMAP5 
  cosmology, measured in concentric spherical shells around void centers.  We 
  estimate the profiles using the average density of all galaxies, measured from 
  the centers of voids.  The radii of the spherical shells are normalized to 
  each void's effective radius, $R_\text{eff}$.  $1 + \delta = 1$ corresponds to 
  the mean survey density, while $1 + \delta = 0$ corresponds to a density of 
  0.}
  \label{fig:vdel2}
\end{figure*}

Due to the range of void sizes in the population, the overdense wall feature in 
the density profiles of smaller voids can overlap with the underdense central 
regions of larger voids, resulting in a smearing of the wall feature in the void 
density profile.  To counteract this effect, we compute a normalized stacked 
void density profile, where the radial bins are scaled in relation to each 
void's effective radius; the resulting density profiles are shown in 
Figure~\ref{fig:vdel2}.  There is a clear enhancement in the strength of the 
central underdensity relative to the wall feature, particularly for the void 
catalogs of VoidFinder and \REVOLVER.

For each of our void catalogs, we find that the mock galaxy catalogs exhibit an 
overdensity in the voids relative to that observed in SDSS~DR7.  This indicates 
that the Horizon Run 4 simulations either use a different galaxy bias than 
exists in the data or produce a matter distribution whose underdense regions are 
not as underdense as the observed galaxy distribution.  If either of these are 
the case, then the mock voids are less empty than the observed voids, so we 
would expect the VoidFinder algorithm to grow slightly smaller voids than in 
data, and to observe slightly more void galaxies in the mock galaxy catalogs 
than in data.  Likewise, we would expect the \VV algorithm to find slightly more 
void galaxies and slightly smaller voids in the mock galaxy catalog than in 
data, since the voids will spread to smaller radii when the density gradient is 
more shallow but the linking density remains fixed.  This is consistent with the 
mock void catalog properties summarized in Tables~\ref{tab:data_table} and 
\ref{tab:overlap}.

For all algorithms used, the density profiles display qualitative agreement with 
previous void studies, both on data \citep[e.g.,][]{Pan12, Ceccarelli13, 
Nadathur14, Ricciardelli14} and simulations \citep[e.g.,][]{Colberg08, Lavaux12, 
Hamaus14b, Sutter14a}.  In particular, void density profiles produced 
specifically from the SDSS~DR7 such as those in \cite{Pan12} and 
\cite{Sutter12a} display a similar behavior at large $r$, where $1+\delta$ 
levels off below $1$ due to the relatively small size of the survey volume.

\begin{deluxetable*}{lll|CCCCC}
  \tablewidth{0pt}
  \tablecaption{Properties of void catalogs\label{tab:data_table}}
  \tablehead{\colhead{} \vspace{-0.2cm} & \colhead{} & \colhead{} & \colhead{} & \colhead{median $R_\text{eff}$} & \colhead{maximum $R_\text{eff}$} & \colhead{\% survey volume} & \colhead{\% galaxies} \\ \colhead{} & \colhead{Cosmology} & \colhead{Algorithm} & \colhead{$N_\text{void}$} & \colhead{[\hMpc]} & \colhead{[\hMpc]} & \colhead{in voids} & \colhead{in voids}}
  \startdata
    \multirow{6}{*}{Data}
    & \multirow{3}{*}{Planck 2018}
    & VoidFinder  & 1163 & 15.5\pm 0.1 & 28.7 & 60.8 & 17.6 \\ 
    & & \VIDE     & 531  & 16.8\pm 0.5 & 53.1 & 68.1 & 64.4\\
    & & \REVOLVER & 518  & 19.4\pm 0.4 & 40.9 & 90.8 & 82.3 \\ 
    \cline{2-8}
    & \multirow{3}{*}{WMAP5}
    & VoidFinder  & 1184 & 15.6\pm 0.1 & 28.7 & 61.1 & 18.2 \\ 
    & & \VIDE     & 534  & 17.1\pm 0.5 & 53.3 & 63.0 & 64.8\\
    & & \REVOLVER & 518  & 19.5\pm 0.4 & 41.2 & 86.5 & 82.3 \\ 
    \hline
    \multirow{3}{*}{Simulations} & \multirow{3}{*}{WMAP5}
    & VoidFinder  & 1283\pm 21 & 15.4\pm 0.1 & 29.3\pm 1.6 & 60.4\pm 0.7 & 18.2\pm 0.4 \\ 
    & & \VIDE     & 558\pm 25  & 16.7\pm 0.4 & 52.4\pm 8.7 & 66.6\pm 3.0 & 62.8\pm 2.6 \\ 
    & & \REVOLVER & 545\pm 19  & 19.2\pm 0.4 & 41.1\pm 3.1 & 88.7\pm 0.5 & 80.4\pm 0.5
  \enddata
  \tablecomments{Properties of void catalogs made from the SDSS~DR7 (top) and 64 
  mock galaxy catalogs created from the Horizon Run 4 simulation (bottom).  The 
  volume-limited catalogs are made with a redshift cut of $z < 0.114$ and an 
  absolute magnitude cut of $M_r < -20$.  Edge voids are excluded from all 
  properties except $N_\text{void}$.}
\end{deluxetable*}

\begin{deluxetable*}{llcc|CC}
  \tablewidth{0pt}
  \tablecaption{Void overlap between catalogs\label{tab:overlap}}
  \tablehead{\vspace{-0.2cm} & & \multicolumn{2}{c|}{Algorithms} & \colhead{\% survey volume} & \colhead{\% galaxies}
  \\
  \vspace{-0.2cm} & & \multirow{2}{*}{$A$} & \multirow{2}{*}{$B$} & \colhead{in voids} & \colhead{in voids}
  \\
  & \colhead{Cosmology} & & & \colhead{($A \cap B$)} & \colhead{($A \cap B$)}
  }
  \startdata
    \multirow{6}{*}{Data}
    & \multirow{3}{*}{Planck 2018}
      & VoidFinder & \VIDE     & 40.9 & 11.0\\
    & & VoidFinder & \REVOLVER & 56.3 & 13.7\\
    & & \VIDE      & \REVOLVER & 63.3 & 58.3\\
    \cline{2-6}
    & \multirow{3}{*}{WMAP5}
      & VoidFinder & \VIDE     & 37.9 & 11.4\\
    & & VoidFinder & \REVOLVER & 51.8 & 14.1\\
    & & \VIDE      & \REVOLVER & 62.9 & 58.6\\
    \hline
    \multirow{3}{*}{Simulations}
    & \multirow{3}{*}{WMAP5}
      & VoidFinder & \VIDE     & 40.0\pm 1.9 & 11.7\pm 0.5 \\
    & & VoidFinder & \REVOLVER & 54.9\pm 0.8 & 15.7\pm 0.4 \\
    & & \VIDE      & \REVOLVER & 62.0\pm 3.0 & 56.8\pm 2.6 \\
  \enddata
  \tablecomments{Percentage of overlap in survey volume and galaxies found in 
  voids in the SDSS DR7 between each of the void catalogs.  Edge voids are 
  excluded from all percentages.}
\end{deluxetable*}

\vspace{-0.75in}

\section{Comparison to previous void catalogs}\label{sec:discussion}

\begin{figure*}
  \centering
  
  \begin{annotationimage}{width=0.485\textwidth}{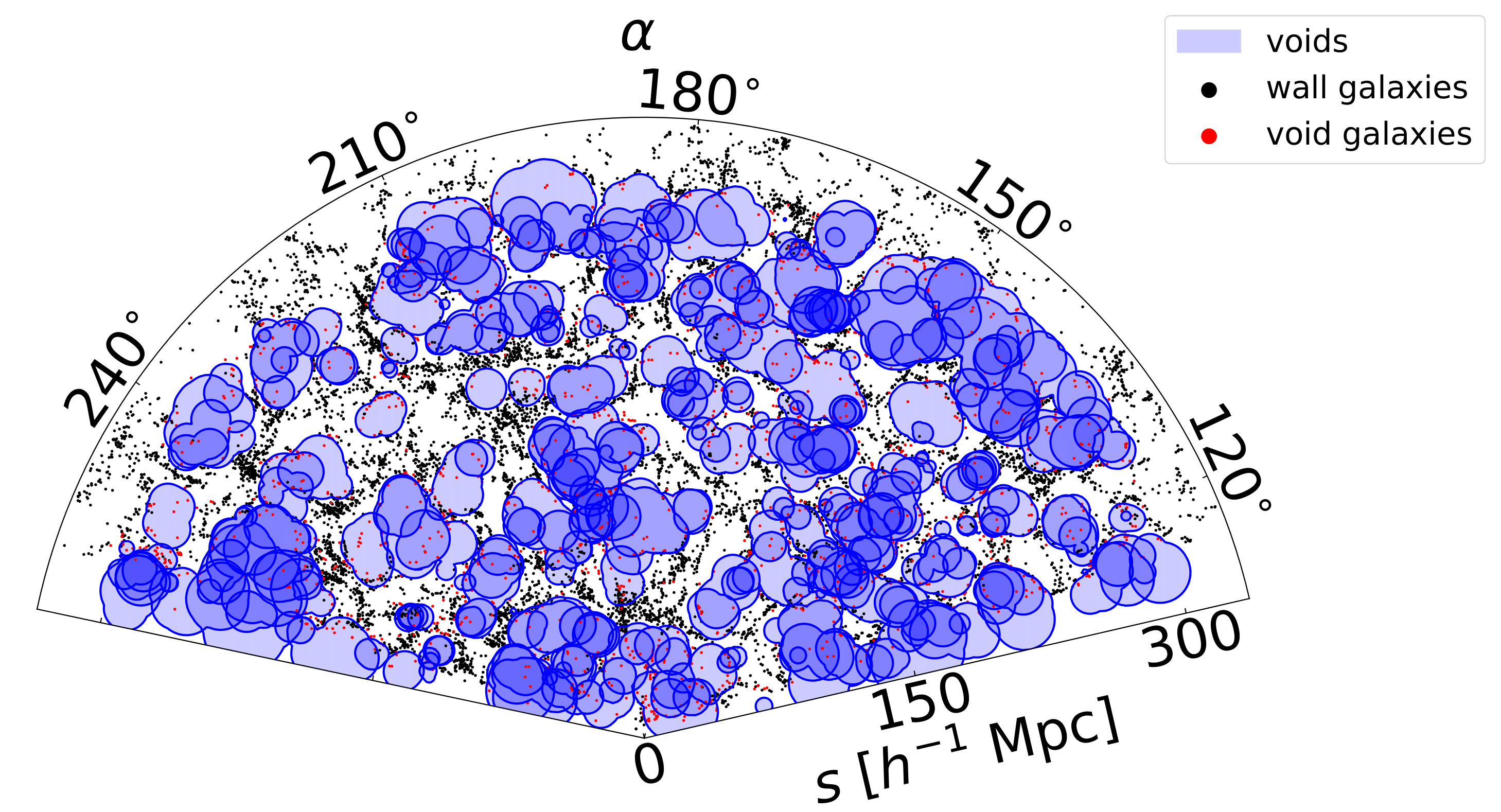}
    \draw[image label = {\cite{Pan12} at north west}];
  \end{annotationimage}
  \begin{annotationimage}{width=0.485\textwidth}{f1a.png}
    \draw[image label = {This work at north west}];
  \end{annotationimage}
  
  \caption{Same region of the SDSS DR7 survey, showing the intersection of the 
  voids published by \cite{Pan12} (left) and our VoidFinder catalog (right) 
  generated assuming a Planck 2018 cosmology.  The excess void regions 
  identified by \cite{Pan12} are easily seen along the boundaries of the survey.  
  Within the footprint, we also see that the void catalog published by 
  \cite{Pan12} has significant overlap between identified voids.}
  \label{fig:VF_comp}
\end{figure*}

Previous void catalogs based on SDSS~DR7 have been created, including 
\cite{Pan12} with VoidFinder, and \cite{Sutter12a} and \cite{Nadathur14} with 
ZOBOV.  Each of these void catalogs is based on a different volume-limited 
version of SDSS~DR7, and each handles the survey boundaries differently.  Our 
void catalogs address these issues and improve upon these earlier catalogs.  
Below, we focus on the improvements in our VoidFinder SDSS~DR7 void catalog 
relative to earlier versions, as comparisons to all previously published void 
catalogs is beyond the scope of this work.

\subsection{Choice of galaxy catalog}

The VoidFinder void catalog described in \cite{Pan12} is based on the Korea 
Institute for Advanced Study Value-Added Galaxy Catalog 
\citep[KIAS-VAGC;][]{Choi10} and its main source the New York University 
Value-Added Galaxy Catalog \citep[NYU-VAGC;][]{Blanton05a}.  The volume-limited 
catalog used by \cite{Pan12} requires galaxies to have $M_r < -20.09$ and 
$z < 0.107$, while we require $M_r < -20.0$ and $z < 0.114$.  The redshift limit 
that we use is better matched to the maximum redshift at which a galaxy with 
$M_r = -20.0$ is observed in the magnitude-limited SDSS~DR7 spectroscopic survey 
($m_r < 17.77$).

We choose to use the NSA as our galaxy catalog for three main reasons.  First, 
the KIAS-VAGC assigns a redshift to galaxies with $m_r < 17.77$ that were not 
spectroscopically targeted in SDSS~DR7 due to their proximity with another 
spectroscopic target (avoiding fiber collisions).  While it is sometimes a safe 
assumption that two galaxies with this small sky separation are actually close 
in redshift, some of these objects have since been spectroscopically observed 
and found to have a significantly different redshift than their projected 
nearest neighbor.  Second, the NSA better handles large galaxies, preventing 
them from being segmented into a number of spurious small unique objects.  
Finally, the NSA also does not include galaxies close to bright stars, whose 
photometry might be affected by the star's light.  All of these properties 
impact the volume-limited catalog used for void-finding, slightly affecting the 
voids found with the algorithms.

\subsection{Handling survey boundaries}

Relative to \cite{Pan12}, our implementation of VoidFinder improves on the 
handling of the survey boundaries and on the merging of smaller spheres into the 
voids.  As seen in Figure~\ref{fig:VF_comp}, the void catalog from \cite{Pan12} 
includes a large number of voids near the survey edge, resulting from the growth 
of spheres into the empty space beyond the survey limits.  This is the source of 
the erroneous large voids included in the void catalog by \cite{Pan12}.  As 
described in Section~\ref{sec:VF}, we mitigate this effect by removing all 
spheres which extend beyond the survey bounds by more than 10\% of their volume.  
Further, there is significant overlap between many of the voids, double-counting 
some of the survey volume and returning voids which are not dynamically 
distinct.  Our implementation of VoidFinder avoids this by succinctly limiting 
the extent to which unique voids can overlap (see Section~\ref{sec:VF} for 
details).

Another way to handle survey boundaries is to populate the edges with an 
artificially high density of galaxies, thereby creating a dense wall through 
which no void will reach.  In this manner, \cite{Sutter12a} placed random 
particles along the survey boundaries to determine which galaxies' Voronoi cells 
should be assigned infinite density to prevent the growth of voids beyond the 
survey edges.  However, the density of these randomly distributed particles is 
not high enough to constrain the Voronoi volumes of the real galaxies, causing 
the central densities of the voids along the boundaries to be lower than 
expected.  Instead, as described in Section~\ref{sec:V2}, \VV sets the volume of 
any Voronoi cell with a vertex outside of the survey boundaries to zero, 
naturally limiting the void growth along the survey edges without generating an 
artificial wall surrounding the survey.

\subsection{Consistent volume-limited galaxy catalog}

Void catalogs previously published for the SDSS~DR7 using different void-finding 
algorithms used different cosmologies as well as different magnitude and 
redshift limits, making a direct comparison of the algorithms' results 
difficult.  Our void catalogs are all created from the same volume-limited 
sample using a single cosmology and with similar void-finding parameters when 
possible (such as the minimum void radius). This allows for simple, direct 
comparison of the results from the various algorithms.

\section{Conclusions}\label{sec:conc}

Using the SDSS DR7 NSA galaxy catalog, we have created a volume-limited catalog 
of 194,125 galaxies with $M_r < -20$ at $z < 0.114$.  We use Python 
implementations of the VoidFinder \citep{ElAd97, Hoyle02} and ZOBOV 
\citep{Neyrinck08} algorithms from the Void Analysis Software Toolkit 
\citep[VAST;][]{Douglass22} to find voids in this volume-limited galaxy catalog.  
We also apply the same void-finding algorithms to 64 mock galaxy catalogs 
produced from the Horizon Run 4 simulation \citep{Kim15}.  Three void catalogs 
based on VoidFinder, \VIDE, and \REVOLVER have been made public\footnote{The 
catalogs are available in \texttt{ASCII} format at 
\href{https://doi.org/10.5281/zenodo.5834786}{https://doi.org/10.5281/zenodo.5834786}.}.

The computed properties of these void populations display good agreement between 
the data and mock catalogs, although \VV tends to find fewer and larger voids 
than VoidFinder.  This is especially true for \VIDE and is a result of the 
linking of underdense ``zones.''  Our findings qualitatively agree with other 
public void catalogs \citep{Pan12, Sutter12a, Nadathur14}, exhibiting a central 
underdensity in the stacked void profiles that is surrounded by an overdensity 
at $r\approx R_\text{eff}$ in the normalized profiles.

Our void catalogs are suited for void-based studies in both cosmology and 
astrophysics.  The positions of voids can be used in void-galaxy 
cross-correlations, in Alcock-Paczy{\'n}ski tests, in studies of CMB lensing by 
voids, and in studies of void ellipticities.  Further, the identification of 
galaxies in voids can be used to study galactic properties in dense vs. 
underdense regions.

Both methods presented here come with particular advantages and disadvantages.  
The watershed technique (\VV) is simple and relatively free of tunable 
parameters, while VoidFinder requires the user to define several parameters 
(e.g., void sphere overlap fraction and field galaxy criterion).  We find that 
\VV includes the shell-crossing regions as part of the void volume, while 
VoidFinder excludes them.  The choice of which void catalog to use then depends 
on the intended science, since the dynamics in the shell-crossing regions around 
voids are significantly different than within the void interiors.  As a result, 
the VoidFinder catalog is especially suited for studies of galaxy properties 
(Zaidouni et al., in prep).  We find the extension of \VV voids into overdense 
regions also leads to a very large fraction of galaxies classified as void.  
Similarly, our void catalogs can be used to study the properties of the 
intergalactic medium --- for example, in searches for the prevalence of 
Lyman-$\alpha$ absorption in voids \citep{Watson22}.

\section*{Acknowledgements}

The authors would like to thank Stephen W. O'Neill, Jr. for his help improving 
the VoidFinder algorithm in VAST, and Michael Vogeley and Fiona Hoyle for the 
original VoidFinder code and void-finding expertise.  D.V. and S.B. acknowledge 
support from the U.S. Department of Energy Office of High Energy Physics under 
the grant DE-SC0008475.  K.D. and D.V. acknowledge support from grant 62177 from 
the John Templeton Foundation.

The authors thank the Center for Integrated Research Computing (CIRC) at the 
University of Rochester for providing computational resources and technical 
support.
  
Funding for the SDSS and SDSS-II has been provided by the Alfred P. Sloan 
Foundation, the Participating Institutions, the National Science Foundation, the 
U.S. Department of Energy, the National Aeronautics and Space Administration, 
the Japanese Monbukagakusho, the Max Planck Society, and the Higher Education 
Funding Council for England.  The SDSS web site is \url{http://www.sdss.org/}.

The SDSS is managed by the Astrophysical Research Consortium for the 
Participating Institutions.  The Participating Institutions are the American 
Museum of Natural History, Astrophysical Institute Potsdam, University of Basel, 
University of Cambridge, Case Western Reserve University, University of Chicago, 
Drexel University, Fermilab, the Institute for Advanced Study, the Japan 
Participation Group, Johns Hopkins University, the Joint Institute for Nuclear 
Astrophysics, the Kavli Institute for Particle Astrophysics and Cosmology, the 
Korean Scientist Group, the Chinese Academy of Sciences (LAMOST), Los Alamos 
National Laboratory, the Max-Planck-Institute for Astronomy (MPIA), the 
Max-Planck-Institute for Astrophysics (MPA), New Mexico State University, Ohio 
State University, University of Pittsburgh, University of Portsmouth, Princeton 
University, the United States Naval Observatory, and the University of 
Washington.

\bibliographystyle{aasjournal}
\bibliography{Doug1122_sources}

\end{document}

%% file: VF_maximals.tex
\begin{deluxetable*}{CRCCCCCCCC}
  \tablewidth{0pt}
  \colnumbers
  \tablecaption{\label{tab:vffile1}VoidFinder output: Maximal spheres}
  \tablehead{\colhead{x} & \colhead{y} & \colhead{z} & \colhead{radius} & \colhead{flag} & \colhead{edge} & \colhead{s} & \colhead{ra} & \colhead{dec} & \colhead{Reff} \\ \colhead{\hMpc} & \colhead{\hMpc} & \colhead{\hMpc} & \colhead{\hMpc} & \colhead{} & \colhead{} & \colhead{\hMpc} & \colhead{deg} & \colhead{deg} & \colhead{\hMpc}}
  \startdata
  -169.041 & 162.884 & 209.440 & 22.441 & 0 & 1 & 314.597 & 136.062 & 41.739 & 27.855 \\ 
  -135.968 &  52.528 & 271.538 & 22.335 & 1 & 1 & 308.187 & 158.877 & 61.772 & 30.611 \\ 
  -167.611 &  95.791 & 251.200 & 21.814 & 2 & 1 & 316.814 & 150.251 & 52.456 & 27.942 \\ 
  -269.456 & -80.652 & 134.977 & 21.761 & 3 & 1 & 311.978 & 196.663 & 25.635 & 29.541 \\ 
  -229.344 & 110.975 & 103.677 & 21.144 & 4 & 0 & 275.069 & 154.178 & 22.142 & 28.694
  \enddata
  \tablecomments{Five of the 1163 maximal spheres of the VoidFinder SDSS~DR7 
  void catalog assuming a Planck 2018 cosmology.  This table is published in its 
  entirety online in a machine-readable format.  A portion is shown here for 
  guidance regarding its form and content.  Columns include: (1--3) Cartesian 
  coordinates of maximal sphere center; (4) Radius of maximal sphere; (5) Unique 
  identifier; (6) Flag identifying whether any part of the void falls outside of 
  the survey mask; (7) Comoving distance of maximal sphere center; (8--9) Sky 
  coordinates of maximal sphere center; (10) Effective radius of the void to 
  which the maximal sphere belongs.}
\end{deluxetable*}

%% file: VF_holes.tex
\begin{deluxetable}{RRRRR}
  \tablewidth{0pt}
  \colnumbers
  \tablecaption{\label{tab:vffile2}VoidFinder output: All spheres}
  \tablehead{\colhead{x} & \colhead{y} & \colhead{z} & \colhead{radius} & \colhead{flag} \\ \colhead{\hMpc} & \colhead{\hMpc} & \colhead{\hMpc} & \colhead{\hMpc} & \colhead{}}
  \startdata
  -169.041 & 162.884 & 209.440 & 22.441 & 0 \\ 
  -169.267 & 163.627 & 210.224 & 22.362 & 0 \\ 
  -135.968 &  52.528 & 271.538 & 22.335 & 1 \\ 
  -169.634 & 162.112 & 207.992 & 22.291 & 0 \\ 
  -136.764 &  52.442 & 271.932 & 22.233 & 1 
  \enddata
  \tablecomments{Five of the 39,735 spheres in the VoidFinder SDSS~DR7 void 
  catalog assuming a Planck 2018 cosmology.  This table is published in its 
  entirety online in a machine-readable format.  A portion is shown here for 
  guidance regarding its form and content.  Columns include: (1--3) Cartesian 
  coordinates of sphere center; (4) Radius of sphere; (5) Unique identifier.  
  The union of all spheres with the same unique identifier form one void.}
\end{deluxetable}

%% file: V2_voids_skinny.tex
\begin{deluxetable*}{RRRRRRRRRRRR}
  \tablewidth{0pt}
  \tablecaption{\label{tab:v2file1}\VV output: Voids}
  \tablehead{\colhead{x} & \colhead{y} & \colhead{z} & \colhead{redshift} & \colhead{ra} & \colhead{dec} & \colhead{radius} & \colhead{x1~\ldots} & \colhead{x2~\ldots} & \colhead{x3~\ldots} & \colhead{area} & \colhead{edge} \\ \colhead{\hMpc} & \colhead{\hMpc} & \colhead{\hMpc} & \colhead{} & \colhead{deg} & \colhead{deg} & \colhead{\hMpc} & \colhead{\hMpc} & \colhead{\hMpc} & \colhead{\hMpc} & \colhead{(\hMpc)$^2$} & \colhead{(\hMpc)$^2$} \\ \colhead{(1)} & \colhead{(2)} & \colhead{(3)} & \colhead{(4)} & \colhead{(5)} & \colhead{(6)} & \colhead{(7)} & \colhead{(8--10)} & \colhead{(11--13)} & \colhead{(14--16)} & \colhead{(17)} & \colhead{(18)}}
  \startdata
  -133.845 & 80.947 & 256.018 & 0.102 & 148.835 & 58.576 & 23.904 & -4.244 & 25.889 & -21.679 & 20096.9 & 2010.5 \\ 
  -245.112 & 134.208 & 113.007 & 0.103 & 151.297 & 22.018 & 26.164 & -8.689 & 22.479 & -6.785 & 25769.4 & 1736.4 \\ 
  -134.931 & 27.948 & 261.080 & 0.100 & 168.297 & 62.175 & 22.857 & -4.066 & 24.292 & 21.974 & 18592.4 & 2719.4 \\ 
  -246.946 & -59.461 & 82.552 & 0.091 & 193.538 & 18.004 & 24.167 & -10.082 & -13.396 & -0.247 & 20203.4 & 0.0 \\ 
  -220.533 & 184.126 & 89.876 & 0.102 & 140.140 & 17.371 & 23.049 & 6.221 & -23.418 & -11.415 & 17058.4 & 1160.7 
  \enddata
  \tablecomments{Five of the 531 voids in the \VV/VIDE SDSS~DR7 void catalog 
  assuming a Planck 2018 cosmology.  This table is published in its entirety 
  online in a machine-readable format; a portion is shown here for guidance 
  regarding its form and content.  Columns include: (1--3) Cartesian coordinates 
  of the void's weighted center; (4) Redshift of the void's weighted center; 
  (5--6) Sky coordinates of the void's weighted center; (7) Effective radius of 
  the void; (8--10) Cartesian components of the void's first ellipsoid axis 
  (only $x$-value is shown); (11--13) Cartesian components of the void's second 
  ellipsoid axis (only $x$-value is shown); (14--16) Cartesian components of the 
  void's third ellipsoid axis (only $x$-value is shown); (17) Total 
  surface area of the void; (18) Surface area of the void which is adjacent to 
  boundary cells.}
\end{deluxetable*}

%% file: V2_zones.tex
\begin{deluxetable}{ccc}
  \tablewidth{0pt}
  \tablecaption{\label{tab:v2file2}\VV output: Zones}
  \tablehead{\colhead{\hspace{0.25in}zone}\hspace{0.25in} & \colhead{\hspace{0.25in}void0}\hspace{0.25in} & \colhead{\hspace{0.25in}void1}\hspace{0.25in} \\ \colhead{(1)} & \colhead{(2)} & \colhead{(3)}}
  \startdata
  0 & -1 & -1 \\
  1 & -1 & -1 \\
  2 & -1 & -1 \\
  3 & 96 & 96 \\
  4 & -1 & -1 
  \enddata
  \tablecomments{Five of the 1037 zones in the \VV/VIDE SDSS~DR7 void catalog 
  assuming a Planck 2018 cosmology.  This table is published in its entirety 
  online in a machine-readable format; a portion is shown here for guidance 
  regarding its form and content.  Columns include: (1) Unique zone identifier; 
  (2) Unique identifier of the smallest void to which the zone belongs; (3) 
  Unique identifier of the largest void to which the zone belongs.  A void's 
  unique identifier is equal to its row number in the file described in 
  Table~\ref{tab:v2file1}.  Values of -1 indicate that the zone is not part of 
  any void.}
\end{deluxetable}

%% file: V2_galaxies.tex
\begin{deluxetable}{ccccc}
  \tablewidth{0pt}
  \tablecaption{\label{tab:v2file3}\VV output: Galaxies}
  \tablehead{\colhead{\hspace{0.1in}gal}\hspace{0.1in} & \colhead{\hspace{0.1in}zone}\hspace{0.1in} & \colhead{\hspace{0.1in}depth}\hspace{0.1in} & \colhead{\hspace{0.1in}edge}\hspace{0.1in} & \colhead{\hspace{0.1in}out}\hspace{0.1in} \\ \colhead{(1)} & \colhead{(2)} & \colhead{(3)} & \colhead{(4)} & \colhead{(5)}}
  \startdata
  3 & 1036 & 0 & 1 & 0 \\ 
  7 & 1036 & 0 & 1 & 0 \\
  10 & 433 & 3 & 0 & 0 \\
  12 & 1036 & 0 & 1 & 0 \\
  13 & 1036 & 0 & 1 & 0 
  \enddata
  \tablecomments{Five of the 194,125 galaxies in the volume-limited SDSS~DR7 
  catalog with void information corresponding to the \VV/VIDE void catalog 
  assuming a Planck 2018 cosmology.  This table is published in its entirety 
  online in a machine-readable format; a portion is shown here for guidance 
  regarding its form and content.  Columns include: (1) Galaxy unique identifier 
  (taken from the input catalog when possible); (2) Unique identifier for the 
  zone to which the galaxy belongs; (3) Number of adjacent Voronoi cells between 
  the galaxy's Voronoi cell and the edge of the zone to which it belongs; (4) 
  Boolean value determining whether or not the galaxy is at the edge of the 
  survey mask; (5) Boolean value determining whether or not the galaxy is 
  outside the survey mask.}
\end{deluxetable}